\documentclass[12pt]{article}

\usepackage{color}

\definecolor{dgreyblue}{rgb}{0.26,0.3,0.46}             

\newcommand{\truered}[1]{{\textcolor{red}{#1}}}

\usepackage{psfig,pifont,xspace}
\usepackage{graphics}
\usepackage{graphicx}      
\usepackage{amssymb}
\usepackage{epsf}
\usepackage{citesort}
\usepackage{amssymb}

\newtheorem{Theorem}{Theorem}
\newtheorem{Lemma}[Theorem]{Lemma}
\newtheorem{Corollary}[Theorem]{Corollary}

\newtheorem{Definition}[Theorem]{Definition}
\newtheorem{proposition}{Proposition}

\newenvironment{proof}{{\noindent\bf Proof.\ }}{\hfill{\Pisymbol{pzd}{113}}\vspace{0.1in}}

\newtheorem{remark}{Remark}
\newtheorem{problem}{Problem}

\newcommand{\TB}{\vspace{-0.1ex}}\newcommand{\TiE}{\setlength{\itemsep}{-1ex}}

\newcommand{\IE}{{\em i.e.}\xspace}

\newcommand{\ignore}[1]{}

\newcommand{\halmos}{\rule{1ex}{1.4ex}}
\newcommand{\qed}{\hfill \halmos} 
\newcommand{\text}[1]{\hbox{\rm \ #1\ \/}}
\newcommand{\be}[1]{\begin{equation}\label{#1}}
\newcommand{\ee}{\end{equation}}
\newcommand{\bi}{\begin{itemize}}
\newcommand{\ei}{\end{itemize}}
\newcommand{\ben}{\begin{enumerate}}
\newcommand{\een}{\end{enumerate}}

\newcommand{\R}{{\mathbb R}}  

\newcommand{\bl}[1]{\begin{Lema}\label{#1}}
\newcommand{\el}{\qed\end{Lema}}
\newcommand{\bt}[1]{\begin{Teo}\label{#1}}
\newcommand{\et}{\end{Teo}}
\newcommand{\epr}{\end{proof}}
\newcommand{\bpr}{\begin{proof}}

\newcommand{\beqn}{\begin{eqnarray*}}
\newcommand{\eeqn}{\end{eqnarray*}}

\newcommand{\norma}[1]{\ensuremath{\parallel \! \! #1 \! \! \parallel}}

\newcommand{\plus}{\ensuremath{\, \stackrel{+}{\rightarrow} \,\,}}
\newcommand{\minus}{\ensuremath{\, \stackrel{-}{\rightarrow} \,}}

\newcommand{\partition}{f_V}
\newcommand{\hpartition}{\hat{f}_V}

\date{}

\title{
Algorithmic and Complexity Results for Decompositions\\
of Biological Networks into Monotone Subsystems
}

\author{
Bhaskar DasGupta\thanks{Department of Computer Science, University of
Illinois at Chicago, Chicago, IL 60607.
Partly supported by NSF grants CCR-0296041, CCR-0206795, CCR-0208749 and
IIS-0346973.
Email: {\tt $\{$dasgupta,yzhang3$\}$@cs.uic.edu}.
}
\and
German Andres Enciso\thanks{Department of Mathematics, Rutgers University,
New Brunswick, NJ 08903.
Email: {\tt german.enciso@gmail.com},
{\tt sontag@math.rutgers.edu}. 
}
\thanks{
Partly supported by NSF grant CCR-0206789 and Dimacs.
}
\and
Eduardo Sontag$^\dagger$\thanks{Partly supported by NSF grants EIA 0205116 and DMS-0504557.}
\and
Yi Zhang$^\ast$
}

\begin{document}

\maketitle

\begin{abstract}
A useful approach to the mathematical analysis of large-scale biological
networks is based upon their decompositions into monotone dynamical systems.  
This paper deals with two computational problems associated to finding
decompositions which are optimal in an appropriate sense.
In graph-theoretic language, the problems can be recast in terms of
maximal sign-consistent subgraphs.
The theoretical results include polynomial-time approximation algorithms 
as well as constant-ratio inapproximability results.
One of the algorithms, which has a worst-case guarantee of 87.9\% from
optimality, is based on the semidefinite programming 
relaxation approach of Goemans-Williamson~\cite{Goemans:1995:JACM}.
The algorithm was implemented and tested on a Drosophila
segmentation network and an Epidermal Growth Factor Receptor pathway model,
and it was found to perform close to optimally.

\end{abstract}

\section{Introduction}

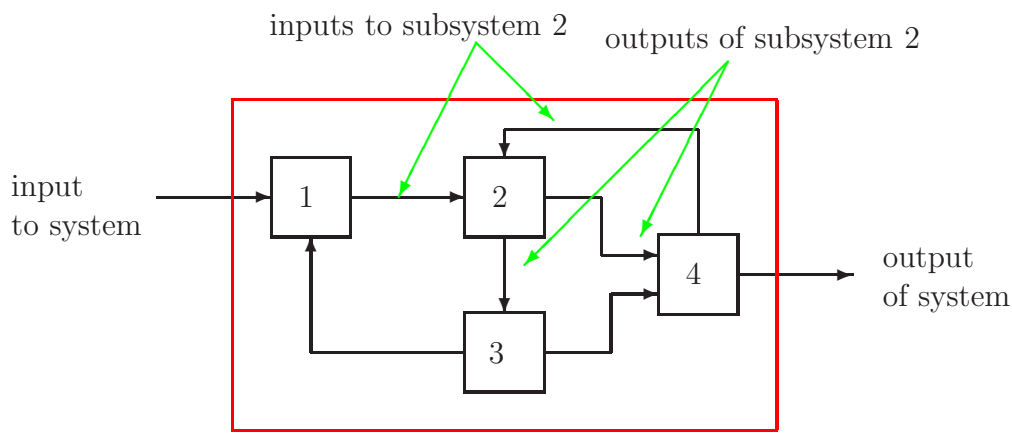
\begin{figure}[h,t]
\begin{center}
\setlength{\unitlength}{3200sp}%
\begin{picture}(5775,3232)(1576,-5183)
\thicklines
\put(2701,-3661){\framebox(600,600){}}
\put(4201,-3661){\framebox(600,600){}}
\put(4201,-4861){\framebox(600,600){}}
\put(5701,-4261){\framebox(600,600){}}
\put(3301,-3361){\vector( 1, 0){900}}
\put(4801,-3361){\line( 1, 0){450}}
\put(5251,-3361){\line( 0,-1){450}}
\put(5251,-3811){\vector( 1, 0){450}}
\put(4801,-4561){\line( 1, 0){525}}
\put(5326,-4561){\line( 0, 1){450}}
\put(5326,-4111){\vector( 1, 0){375}}
\put(4501,-3661){\vector( 0,-1){600}}
\put(4201,-4561){\line(-1, 0){1200}}
\put(3001,-4561){\vector( 0, 1){900}}
\put(6001,-3661){\line( 0, 1){825}}
\put(6001,-2836){\line(-1, 0){1500}}
\put(4501,-2836){\vector( 0,-1){225}}
\put(6301,-3961){\vector( 1, 0){900}}
\put(1801,-3361){\vector( 1, 0){900}}
\thicklines
\truered{\put(2401,-5161){\framebox(4200,2550){}}}
{\color{green}\thicklines
\put(4201,-2161){\vector( 1,-1){600}}
\put(4201,-2161){\vector(-1,-2){600}}
\put(6151,-2311){\vector(-1,-1){1575}}
\put(6151,-2311){\vector(-1,-2){675}}%
\color{black}}%
\put(2825,-3436){1}
\put(4325,-3436){2}
\put(4300,-4650){3}
\put(5825,-4036){4}
\put(5200,-2200){outputs of subsystem 2}
\put(2600,-2100){inputs to subsystem 2}
\put(7351,-3900){output}
\put(7351,-4200){of system}
\put(600,-3350){input}
\put(600,-3650){to system}
\end{picture}
\caption{A system composed of four subsystems}
\label{fig-systems}
\end{center}
\end{figure}

In living cells, networks of proteins, RNA, DNA, metabolites, and other
species process environmental signals, control internal events such as gene
expression, and produce appropriate cellular responses.
The field of systems (molecular) biology is largely concerned with the
study of such networks, viewed as dynamical systems.
One approach to their mathematical analysis relies upon viewing them as
made up of subsystems whose behavior is simpler and easier to understand.
Coupled with appropriate interconnection rules, the hope is that emergent
properties of the complete system can be deduced from the understanding of
these subsystems. 
Diagrammatically, we picture this as in Figure~\ref{fig-systems}, which shows
a full system as composed of four subsystems.

A particularly appealing class of candidates for ``simpler behaved''
subsystems are \emph{monotone systems}, as
in~\cite{Hirsch,Hirsch2,Smithmonotone}.
Monotone systems are a class of dynamical systems for which pathological
behavior (``chaos'') is ruled out.
Even though they may have arbitrarily large dimensionality,
monotone systems behave in many ways like one-dimensional systems.
For instance, in monotone systems, bounded trajectories generically converge
to steady states, and there are no stable oscillatory behaviors.
More precisely, see below, one must extend the notion of monotone system so as
to incorporate input and output channels, as introduced and initially
developed in~\cite{Sontag:mono};
inputs and outputs are required so that interconnections like those shown in
Figure~\ref{fig-systems} can be defined.

Monotonicity is closely related, as explained later, to positive and feedback
loops in systems.  The topic of analyzing the behaviors of such feedback loops
is a long-standing one in biology in the context of regulation, metabolism, and
development; a classical reference in that regard is the
work~\cite{monod} of Monod and Jacob in 1961.
See also, for example,
\cite{lewis77,meinhardt78,thomas78,plathe95,snoussi98,cinquin,angeli-scl-multi,pnas,remi03}.

An interconnection of monotone subsystems, that is to say, an entire
system made up of monotone components, may or may not be monotone: ``positive
feedback'' (in a sense that can be made precise) preserves monotonicity, 
while ``negative feedback'' destroys it. 
Thus, oscillators such as circadian rhythm generators require negative
feedback loops in order for periodic orbits to arise, and hence are
not themselves monotone systems, although they can be decomposed into 
monotone subsystems (cf.~\cite{angeli-circadian-cdc}).
A rich theory is beginning to
arise, characterizing the behavior of non-monotone interconnections.
For example, \cite{Sontag:mono} shows how to preserve convergence to
equilibria; see also the follow-up papers
\cite{leenheer-almost,patrick-lotka,Enciso:Sontag:DCDS2005,halsmithJDE,gedeon}.
Even for monotone interconnections, the decomposition approach is very
useful, as it permits locating and characterizing the stability of
steady states based upon input/output behaviors of components,
as described in~\cite{angeli-scl-multi};
see also the follow-up papers
\cite{pnas,enciso-scl,leenheer-malisoff}.

Moreover, a key point brought up in~\cite{sysbio} is that new techniques for
monotone systems in many situations allow one to characterize the behavior of
an entire system, based upon the ``qualitative'' knowledge represented by
general network topology and the inhibitory or activating character of
interconnections, combined with only a relatively \emph{small amount of
quantitative} data.  The latter data may consist of steady-state responses of
components (dose-response curves and so forth), and there is no need to know
the precise form of dynamics or parameters such as kinetic constants in order
to obtain global stability conclusions.

In Section~\ref{sec-monotone} of this paper, we briefly discuss monotonicity
of systems described by ordinary differential equations (the study of
monotonicity can be extended to partial differential equations,
delay-differential equations, and even more arbitrary dynamical systems, see
e.g.~\cite{Enciso:Sontag:DCDS2005} in the context of monotone systems with
inputs and outputs).
We explain there how the study of monotone systems, and more generally of
decompositions into monotone systems, relates to a \emph{sign-consistency}
property for the graph which describes how each state variable influences each
other variable in a given system.

\begin{figure}[h,t]
\begin{center}
\includegraphics[scale=0.2]{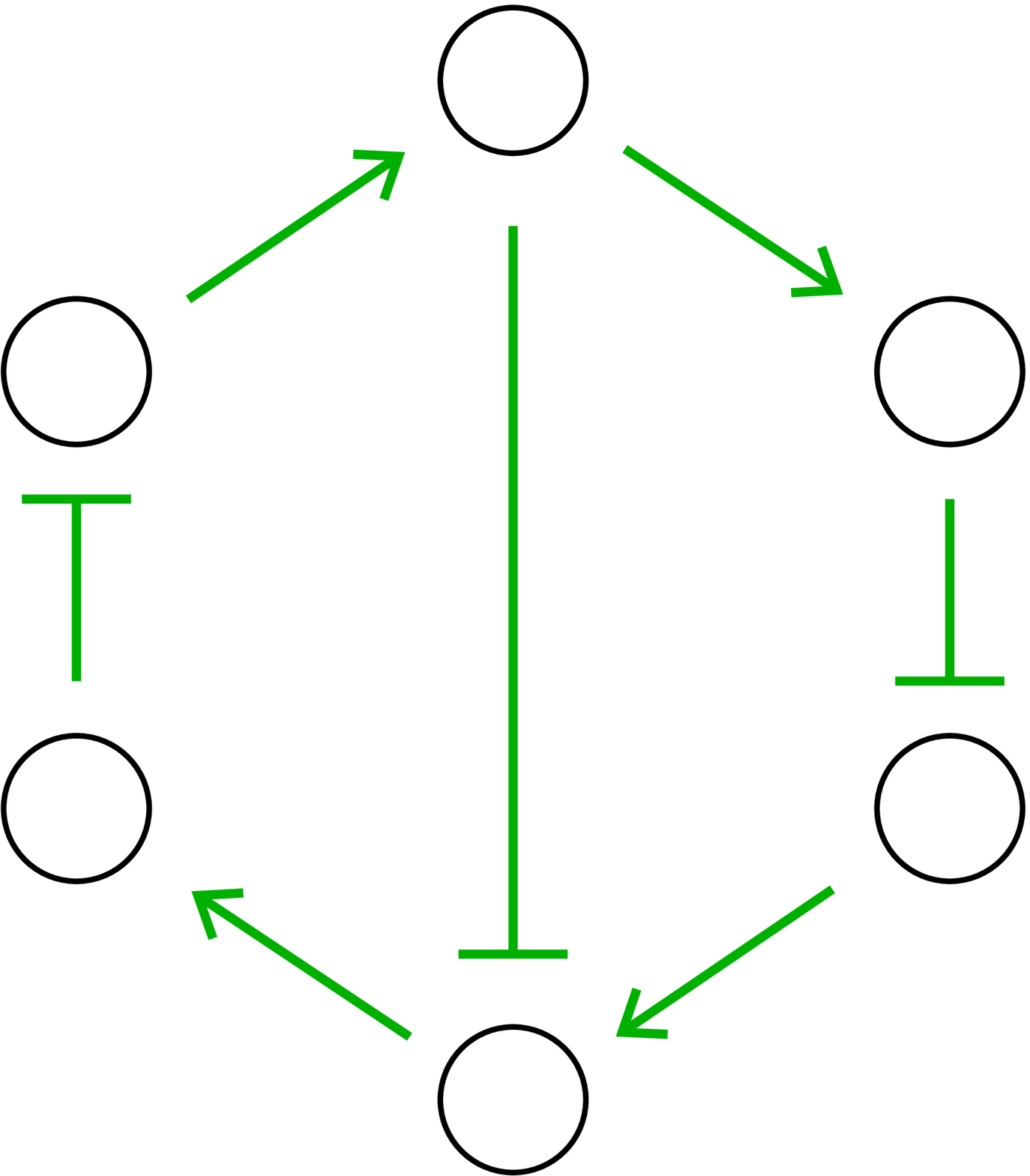}
\ \ \ \ \ \ \ \ 
\includegraphics[scale=0.2]{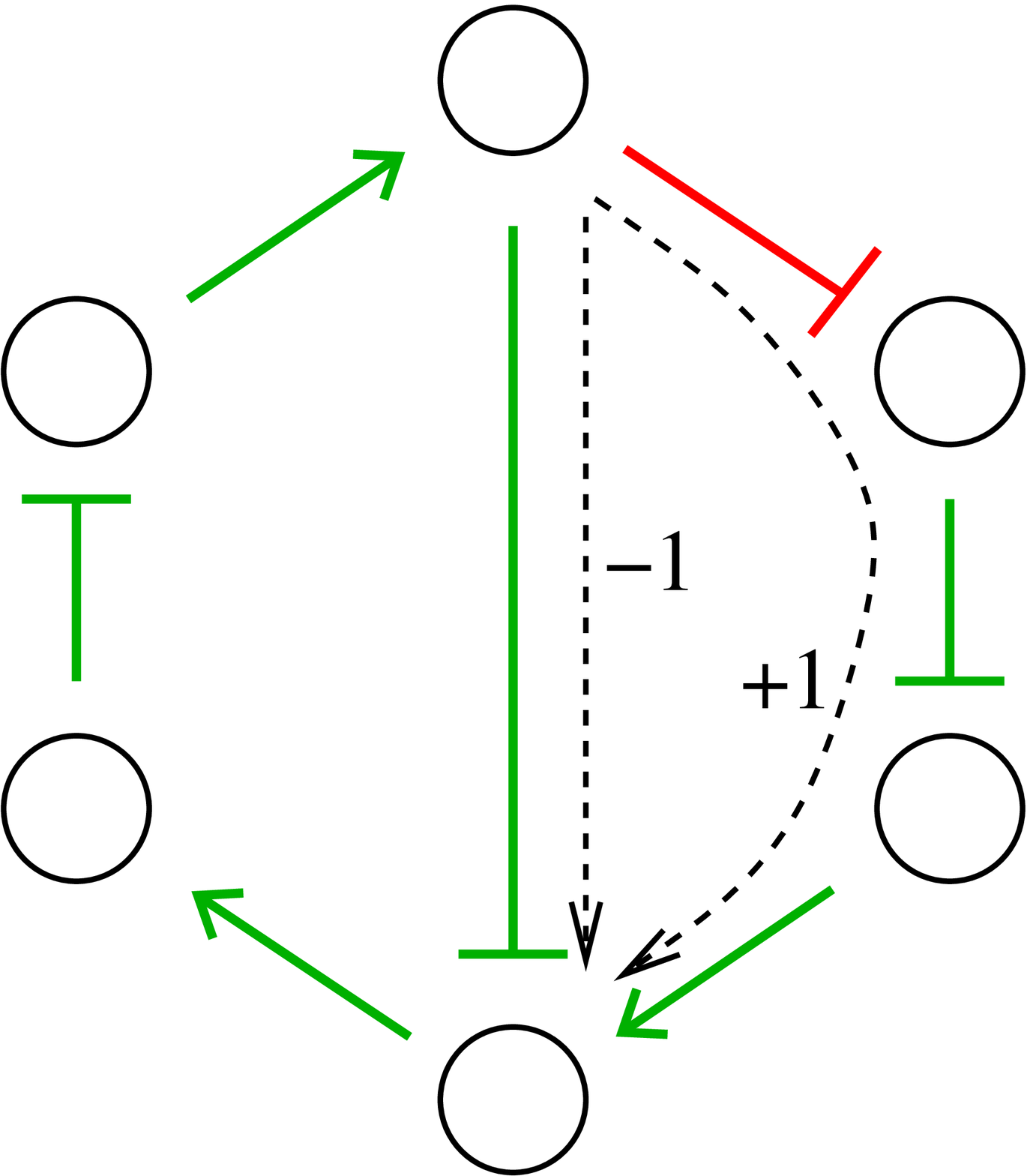}
\caption{A consistent and an inconsistent graph}
\label{one-of-each}
\end{center}
\end{figure}

Generally, a graph, whose edges are labeled by ``$+$'' or ``$-$'' signs
(sometimes one writes $+1,-1$ instead of $+,-$, or uses
respectively activating  ``$\rightarrow$'' or inhibiting
``$\dashv$'' arrows as shown in Figure~\ref{one-of-each}),
is said to be \emph{sign-consistent} if all paths between any two nodes
have the same net sign, or equivalently, all closed loops have positive
parity, i.e.\ an even number, possibly zero, of negative edges.
(For technical reasons, one ignores the direction of arrows, looking only at
undirected graphs; see more details in Section~\ref{sec-monotone}.)
Thus, the first graph in Figure~\ref{one-of-each} is consistent, but the
second one, which differs in just one edge from the first one,
is not (two paths with different parity are shown).

\begin{figure}[h,t]
\begin{center}
\setlength{\unitlength}{1800sp}%
\begin{picture}(3024,1449)(3289,-2998)
\thicklines
\put(3601,-2161){\framebox(2400,600){}}
\put(6001,-1861){\line( 1, 0){300}}
\put(6301,-1861){\line( 0,-1){900}}
\put(6301,-2761){\vector(-1, 0){900}}
\put(3301,-1861){\vector( 1, 0){300}}
\put(3301,-1861){\line( 0,-1){900}}
\put(3301,-2761){\line( 1, 0){900}}
\put(4201,-2986){\framebox(1200,450){}}
\put(4000,-2000){consistent}%
\put(4500,-2900){``$-$''}%
\end{picture}
\caption{Pulling-out inconsistent connections}
\label{pull-out}
\end{center}
\end{figure}
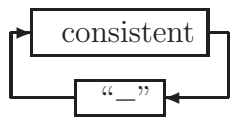

When applying decomposition theorems such as those described in
\cite{Sontag:mono,angeli-scl-multi,pnas,sysbio,leenheer-almost,patrick-lotka,enciso-scl,leenheer-malisoff,Enciso:Sontag:DCDS2005,halsmithJDE,gedeon},
it tends to be the case that
\emph{the fewer the number of interconnections among components, the easier it
  is to obtain useful conclusions.}
One may view a decomposition into interconnections of monotone subsystems
as the ``pulling out'' of ``inconsistent'' connections among 
monotone components, the original system being a ``negative
feedback'' loop around an otherwise consistent system, as represented in
Figure~\ref{pull-out}.
In this interpretation, the number of interconnections among monotone
components corresponds to the number of variables being fed-back.
In addition, and independently from the theory developed in the above
references, one might speculate that nature tends to favor systems that are
decomposable into small monotone interconnections, since ``negative''
feedback loops, although required for homeostasis and for periodic behavior,
have potentially destabilizing effects, especially if there are signal
propagation delays.
Some evidence is provided by work in progress such as
\cite{maayan}, where the authors compare certain biological networks and
appropriately randomized versions of them and show that the original networks
are closer to being consistent,
and by \cite{illya}, where the authors show that, in a Boolean setting,
and using a mean-field calculation of sensitivity, networks of Boolean
functions behave in a more and more ``orderly'' fashion the closer that the
components are to being monotone.

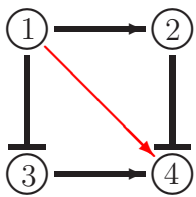
\begin{figure}[h,t]
\begin{center}
\setlength{\unitlength}{3000sp}%
\begin{picture}(1552,1550)(3425,-4436)
\thicklines
\linethickness{2pt}
\put(4801,-3061){\circle{336}}
\put(3601,-4261){\circle{336}}
\put(4801,-4261){\circle{336}}
\put(3601,-3061){\circle{336}}
\put(3751,-3211){\truered{\vector( 1,-1){900}}}
\put(3826,-3061){\vector( 1, 0){750}}
\put(3826,-4261){\vector( 1, 0){750}}
\put(4801,-4036){\line( 0, 1){750}}
\put(3601,-4036){\line( 0, 1){750}}
\put(3451,-4036){\line( 1, 0){300}}
\put(4651,-4036){\line( 1, 0){300}}
\put(4741,-3151){$2$}%
\put(3551,-4361){$3$}%
\put(4731,-4341){$4$}%
\put(3546,-3146){$1$}%
\end{picture}
\caption{Dropping the diagonal edge gives consistency}
\label{5-node-figure}
\end{center}
\end{figure}

Thus, we are led to the subject of this paper, namely computing the smallest
number of edges that have to be removed so that there remains a consistent
graph.
For example, for the particular graph shown in Figure~\ref{5-node-figure}
the answer is that one edge (the diagonal positive one) suffices, and it
is worth remarking that no single other edge would suffice.

In this paper, we will study the computational complexity of the question of
how many edges must be removed in order to obtain consistency, and we provide
a relaxation-based polynomial-time approximation algorithm guaranteed to solve
the problem to about 87.9\% of the optimum solution, which is
based on the semidefinite programming 
relaxation approach of Goemans-Williamson~\cite{Goemans:1995:JACM}
(A variant of the problem is discussed as well).
We also observe that 
it is not possible to have a polynomial-time 
algorithm with performance too close to the optimal.
While our
emphasis is on theory, one of the algorithms was implemented, and we show
results of its application to a Drosophila segmentation network and to an
Epidermal Growth Factor Receptor pathway model.
It turns out that, when applying the algorithm, often the solution is much
closer to optimal than the worst-case guarantee of 87.9\%, and indeed often
gives an optimal solution.

The organization of this paper is as follows.

Section~\ref{sec-monotone} briefly discusses monotonicity.
The discussion is self-contained for the purposes of this paper, and
references are given to the dynamical systems results that motivate the
problem studied here.
The connection to consistency is also explained there.
Section~\ref{Statement of Problem} discusses the associated graph-theoretic
problems and notions of approximability used in the paper,
leading to the statement of our main theoretical results
in Section~\ref{Our Results},
which are proved in Section~\ref{proof of main theorem}.
Section~\ref{appl-sec} contains the mentioned examples of application of
the algorithm.
Several technical proofs are separately provided in an Appendix.

\section{Monotone Systems and Consistency}
\label{sec-monotone}

We will illustrate the motivation for the problem studied here using
systems of ordinary differential equations
\be{system}
\dot{x}=F(x)
\ee
(the dot indicates time derivative, and $x=x(t)$ is a vector), although 
the discussion applies as well to more general types of dynamical systems such
as delay-differential systems or certain systems of reaction-diffusion partial
differential equations.
In applications to biological networks, the component $x_i(t)$ of the
vector $x=x(t)$ indicates the concentration of the $i$th species in the
model at time $t$.
We will restrict attention to models in which the direct effect that one given
variable in the model has over another is either consistently inhibitory or
consistently promoting.  Thus, if protein A binds to the promoter region of
gene B, we assume that it does so either to consistently prevent the
transcription of the gene or to consistently facilitate it. 
(Of course, this condition does not prevent protein A from having an indirect
influence, through other molecules, perhaps dimmers of A itself, that can
ultimately lead to the opposite effect on gene B.)
Mathematically, we require that for every $i,j=1\ldots n$, $i\not=j$, the
partial derivative $\partial F_i / \partial x_j$ be either $\geq0$ at all
states or $\leq 0$ at all states.

Given any partial order $\leq$ defined on $\R^n$, a
system (\ref{system}) is said to be {\em monotone with respect to $\leq$} if
$x_0\leq y_0$ implies $x(t)\leq y(t)$ for every $t\geq0$.  Here $x(t)$, $y(t)$
are the solutions of (\ref{system}) with initial conditions $x_0$, $y_0$,
respectively.  Of course, whether a system is monotone or not depends on the
partial order being considered, but we one says simply that a system
is \emph{monotone} if the order is clear from the context.
Monotonicity with respect to nontrivial orders rules out chaotic attractors
and even stable periodic orbits; see~\cite{Hirsch,Hirsch2,Smithmonotone}, and is, as
discussed in the introduction, a useful property for components when analyzing
larger systems in terms of subsystems.

A useful way to define partial orders in $\R^n$, and the only one to be further
considered in this paper, is as follows.  
Given a tuple $s=(s_1,\ldots s_n)$, where $s_i\in \{1,-1\}$ for every $i$, we
say that $x\leq_s y$ if $s_ix_i\leq s_iy_i$ for every $i$. 
For instance, the ``cooperative order'' is the orthant order 
$\leq_s$ generated by $s=(1,\ldots1)$.   
This is the order
$\leq$  defined by $x\leq y$ if and only if $x_i\leq y_i$ for all
$i=1,\ldots,n$.
It is not difficult to verify if a system is cooperative with respect to
an orthant order; the following lemma, known as ``Kamke's condition,'' is 
not hard to prove, see~\cite{Smithmonotone} for details
(also~\cite{Sontag:mono} in the more general context of monotone systems with
input and output channels).

\begin{Lemma}\label{characterization1} 
Consider an orthant order $\leq_s$ generated by $s=(s_1,\ldots,s_n)$.    
A system (\ref{system}) is monotone with respect to $\leq_s$ if and only if 
\be{char 1 eq}
s_is_j\, \frac{\partial F_j }{ \partial x_i } \geq 0, 
\ i,j=1\ldots n, \ i\not= j.   
\ee
\end{Lemma}

To provide intuition, let us sketch the sufficiency part of the proof for the
special case of the cooperative order.
Suppose by contradiction that the system is not monotone, and that therefore
there is a pair of initial conditions $x_0\leq y_0$ whose solutions $x(t)$,
$y(t)$ cease to satisfy $x(t)\leq y(t)$ at some point.
This implies that at a certain critical moment in time $t$, there is
some coordinate $i$ so that $x_i(t^-)<y_i(t^-)$ but $x_i(t^+)>y_i(t^+)$.
(This argument is not entirely accurate, but it gives the flavor of the
proof.)
Thus $x_i(t)=y_i(t)$ for some $i$ and the derivative with respect to time
of $x_i$ is larger than that of $y_i$ at time $t$, meaning
that that $F_i(x) > F_i(y)$, where $x=x_i(t)$ and $y=y_i(t)$.
However, this cannot happen if $F_i$ is increasing on all the variables $x_j$
except possibly $x_i$, so that $x\leq y,\ x_i=y_i$ implies $F_i(x)\leq F_i(y)$.
An equivalent way to phrase this condition is by ask that
$\partial F_i / \partial x_j \geq 0$ at all states for every $i,j,i\not=j$,
which is the Kamke condition for the special case of the cooperative order.
The name of the order arises because in a monotone system with respect 
to that order each species promotes or ``cooperates'' with each other.  

A rephrasing of this characterization of monotonicity with respect to orthant
orders can be given by looking at the {\em signed digraph} associated to
(\ref{system}) and defined as follows. 
Let $V(G)=\{1,\ldots,n\}$.
Given vertices $i,j$, let $(i,j)\in E(G)$ and $f_E(i,j)=1$ if both 
$\partial F_j / \partial x_i \geq 0$ and the strict inequality holds at least
at one state.
Similarly let $(i,j)\in E(G)$ and $f_E(i,j)=-1$ if 
both $\partial F_j / \partial x_i \leq 0$ and the strict 
inequality holds at least at one state. 
Finally, let $(i,j)\not\in E(G)$ if $\partial F_j / \partial x_i \equiv 0$.  
Recall that we are assuming that one of the three cases must hold.

Now we can define an orthant cone using any function 
$f_V:V(G)\to \{-1,1\}$, by letting $x\leq_{f_V} y$ if and only if 
$f_V(i)x_i \leq f_V(i) y_i$ for all $i$.   
Given $f_V$, we define the consistency function 
$g:E(G)\to \{\mbox{true, false}\}$ by 
$g(i,j)=f_V(i)f_V(j)f_E(i,j)$.  
Then, the following analog of Lemma~\ref{characterization1} holds.

\begin{Lemma}\label{characterization2}
Consider a system (\ref{system}) and an orthant cone $\leq_{f_V}$.  
Then (\ref{system}) is monotone with respect to  $\leq_{f_V}$ if and only if  $g(i,j)\equiv 1$ on $E(G)$.  
\end{Lemma}

\bpr
Let $s_i=f_V(i),\ i=1\ldots n$.  Note that $s_is_j\partial f_i /\partial x_j=0$ if $(i,j)\not\in E(G)$.  
For $(i,j)\in E(G)$, it holds that $s_is_j \partial f_i /\partial x_j\geq0$ if and only if 
$s_is_jf_E(i,j)=1$, that is, if and only if $g(i,j)=1$. The result follows from Lemma~\ref{characterization1}.  
\epr

For the next lemma, let the {\em parity} of a chain in $G$ be the product of
the signs
($+1,-1$) of its individual edges. We will consider in the next result closed 
{\em undirected chains}, that is, sequences $x_{i_1} \ldots x_{i_r}$ such that 
$x_{i_1}=x_{i_r}$, and such that for every 
$\lambda=1,\ldots, r-1$ either $(x_{1,\lambda},x_{1,{\lambda+1}})\in E(G) $ or $(x_{1,{\lambda+1}}, x_{1,\lambda})\in E(G)$.  

The following lemma 
(see~\cite{deangelis} as well as~\cite[page 101]{smith_SIAM88})
is analogous to the fact from vector
calculus that path integrals of a vector field are independent of the
particular path of integration if and only if there exists a potential
function.
Since the result is key to the formulation of the problem being considered,
we provide a simple and self-contained proof in an Appendix.

\begin{Lemma}\label{chainlemma} 
Consider a dynamical system (\ref{system}) with associated directed graph $G$.
Then (\ref{system}) is monotone with respect to some orthant order if and only
if all closed undirected chains of $G$ have parity $1$.   
\end{Lemma}

\subsection{Systems with Inputs and Outputs}

As we discussed in the introduction, a useful approach to the analysis of
biological networks consists of decomposing a given system into an
interconnection of monotone subsystems.
The formulation of the notion of interconnection requires subsystems to
be endowed with ``input and output channels'' through which information
is to be exchanged.
In order to address this we consider {\em controlled} dynamical systems
(\cite{mct}, which
are systems with an additional parameter $u\in \R^m$, and which have the form
\be{controlled system}
\dot{x}=g(x,u).
\ee
The values of $u$ over time are specified by means of a function 
$t\rightarrow u(t)\in \R^m$, $t\geq0$, called an {\em input} pr control.
Thus each input defines a time-dependent dynamical system in the usual sense.
To system (\ref{controlled system}) there is associated a 
{\em feedback function} $h:\R^n\rightarrow \R^m$, 
which is usually used  to create the closed loop system $\dot{x}=g(x,h(x))$.
Finally, if $\R^n,\R^m$ are ordered by orthant orders 
$\leq_{f_V}, \ \leq_q$ respectively, we say that the system is monotone 
if it satisfies (\ref{char 1 eq}) for every $u$, and also 
\be{char 2 eq}
q_k f_V(j) \,\frac{\partial g_j}{\partial u_k} \geq 0,\  
\mbox{for every}\ k,j     
\ee
(see also~\cite{Sontag:mono}.) 
As an example, let us consider the following biological model of testosterone 
dynamics~\cite{Murray2002,Enciso:Sontag:JMB2004}:
\be{testosterone} 
      \begin{array}{l} 
         \dot{x}_1=\frac{\displaystyle A}{\displaystyle K+x_3}-b_1 x_1 \\
         \dot{x}_2=c_1 x_1 - b_2 x_2 \\
         \dot{x}_3=c_2 x_2 - b_3 x_3.   
      \end{array}
\end{equation}
Drawing the digraph of this system, it is easy to see that it is not monotone
with respect to any orthant order, as follows by application of  
Lemma~\ref{chainlemma}.
On the other hand, replacing $x_3$ in the first equation by $u$, we obtain a
system that is monotone with respect to the
orders $\leq_{(1,1,1)},\ \leq_{(-1)}$ for state and input respectively.    
Defining $h(x)=x_3$, the closed loop system of this controlled system is none
other than (\ref{testosterone}). 
The paper~\cite{Enciso:Sontag:JMB2004} shows how, using this decomposition
together with the ``small gain theorem'' from monotone input/output theory
(\cite{Sontag:mono}) leads one to a proof that the system does not have
oscillatory behavior, even under arbitrary delays in the feedback loop,
contrary to the assertion made in~\cite{Murray2002}.

We can carry out this procedure on an arbitrary system (\ref{system}) with a
directed graph $G$, as follows:
given a set $E$ of edges in $G$, enumerate the edges in $E^C$ as
$(i_1,j_1),\ldots (i_m,j_m)$.
For every $k=1\ldots m$, replace all appearances of $x_{i_k}$ in the function
$F_{j_k}$ by the variable $u_k$, to form the function $g(x,u)$.
Define $h(x)=(x_{i_1},\ldots x_{i_m})$.    
It is easy to see that this controlled system (\ref{controlled system}) has
closed loop (\ref{system}).

Note that the controlled system (\ref{controlled system}) generated by the set
$E$ as above has, as associated digraph, 
the sub-digraph of $G$ generated by $E$. 
This is because for every $k$, one has 
$\partial g_{j_k}(x,u) /\partial x_{i_k}\equiv0$, i.e.,\ the edge from $i_k$
to $j_k$ has been ``erased''. 

Let the set $E$ be called {\em consistent} if the undirected subgraph of $G$
generated by $E$ has no closed chains with parity $-1$. 
Note that this is equivalent to the existence of $f_V$ such that $g\equiv 1$
on $E$, by Lemma~\ref{lemmachains} applied to the open loop system
(\ref{controlled system}).
If $E$ is consistent, then the associated system (\ref{controlled system})
itself can also be shown to be monotone:
to verify condition (4), simply define each $q_k$ so that (\ref{char 2 eq}) is
satisfied for $k,j_k$.
Since $\partial g_{j_k}/\partial u_k = \partial F_{j_k}/\partial x_{i_k}\not\equiv 0$, this choice is in fact unambiguous.    
Conversely, if (\ref{controlled system}) is monotone with respect to the
orthant orders $\leq_{f_V},\ \leq_q$, 
then in particular it is monotone for every fixed constant $u$, so that $E$ is consistent by Lemma~\ref{chainlemma}.     
We thus have the following result.    

\begin{Lemma} \label{lemmachains}
Let $E$ be a set of edges of the digraph $G$.    
Then $E$ is consistent if and only if the corresponding controlled system (\ref{controlled system}) 
is monotone with respect to some orthant orders.    
\end{Lemma}

\section{Statement of Problem}
\label{Statement of Problem}

A natural problem is therefore the following.  Given a dynamical system
(\ref{system}) that admits a digraph $G$, use the procedure above to decompose
it as the closed loop of a monotone controlled system (\ref{controlled
system}), while minimizing the number $\norma{E^C}$ of inputs.  Equivalently,
{\em find $f_V$ such that $P(E_+)=\norma{E_+}$ is maximized and
$P(E_-)=\norma{E_-}=\norma{E_+^C}$ minimized}.  This produces the following
problem formulation.

\begin{problem}[Undirected Labeling Problem($ULP$)]\label{prob1}:

\noindent
An instance of this problem is $(G,h)$, where $G=(V,E)$ is an undirected graph and 
$h\colon E\mapsto\{0,1\}$.
A valid solution is a vertex labeling function $f\colon V\rightarrow\{0,1\}$.
Define an edge $\{u,v\}\in E$ to be consistent iff  
$h(u,v)\equiv (f(u)+f(v)) \pmod{2}$.
The objective is then to find a valid solution {\em maximizing} $|F|$ where
$F$ is the set of consistent edges.
\end{problem}

That $ULP$ is a correct formulation for our problem is confirmed by the 
following easy equivalence.

\begin{proposition}
Consider an instance $(G,h)$ of $ULP$ with an optimal solution 
having $x$ consistent edges given by a vertex labeling function $f$. 
Let $D$ be a set of edges of smallest cardinality that have to be removed 
such that for the remaining graph, that is the graph $G'=(V,E\setminus D)$ with 
the same vertex set $V$ but an edge set $E\setminus D$, there exists a vertex
labeling function $f'\colon V\rightarrow \{0,1\}$ that makes every edge consistent.
Then, $x=|E|-|D|$.
\end{proposition}

\begin{proof}
Since $f$ produces a solution of $ULP$ with $x$ consistent edges, exactly 
$|E|-x$ edges are inconsistent, thus $|D|\leq |E|-x$, that is,
$x\leq |E|-|D|$.
Conversely,
since there is a solution with $|E|-|D|$ consistent edges, 
$x\geq |E|-|D|$.
\end{proof}

A special case of $ULP$, namely when  
$h(e)=1$ for all $e\in E$, is the MAX-CUT problem (defined in Section~\ref{approx-section}).
Moreover, $ULP$ can be posed as a special type of 
``constraint satisfaction problem'' as follows.
We have $|E|$ linear equations over $GF(2)$, one equation per edge and each equation
involving exactly two variables, over $|V|$ Boolean variables. The goal is to
assign values to the variables to satisfy the maximum number of equations.
For algorithms and lower-bound results for general cases of these types of 
problems, such as when the equations are over $GF(p)$ for an arbitrary prime
$p>2$, when  
there are an arbitrary number of variables
per equation or when the goal is to minimize the number of unsatisfied equations, see
references such as~\cite{AK96,BK01,CKS01,JV02} and the references therein.

Given orthant orders $\leq_{f_V}$ and $\leq_q$ for $\R^n$ and $\R^m$ respectively, we say that a feedback function 
$h$ is {\em positive} if $x\leq_{f_V} y$ implies $h(x)\leq_q h(y)$, and that it is {\em negative} 
if $x\leq_{f_V} y$ implies $h(x)\geq_q h(y)$.    
It can be shown that the closed loop of a monotone system with a positive feedback function is actually itself 
monotone, so that no system can be produced in this way that was not monotone already.    
But if $h$ is a negative feedback function, then several results become available which use the methods of 
monotone systems for systems that are not monotone, see~\cite{Sontag:mono,Enciso:Sontag:JMB2004,Enciso:Sontag:DCDS2005}.     
For the following result, let $(\mathcal{C},\subseteq)$ be the class of consistent subsets of $E(G)$, ordered under inclusion.

\begin{proposition}\label{propnegativefeedback}
Let $E$ be a consistent set. Then $E$ is maximal in $(\mathcal{C},\subseteq)$ if and only if $h$ is a 
negative feedback function for every $f_V$ such that $g\equiv 1$ on $E$.  
\end{proposition}

\bpr
Suppose that $E$ is maximal, and let $f_V$ be such that $g\equiv 1$ on $E$.    
Given any edge $(i_k,j_k)\in E^C$, it holds that $g(i_k,j_k)=-1$.  
Otherwise one could extend $E$ by adding $(i_k,j_k)$, thus violating maximality.    
That is, $f_V(i_k) f_V(j_k) f_E(i_k,j_k)=-1$.  By monotonicity, it holds that $q_k f_V(j_k) \partial g_{j_k}/\partial u_k \geq0$, 
and since $\partial g_{j_k}/\partial u_k =\partial F_{j_k}/\partial x_{i_k}$, 
it follows necessarily that 
\[
q_k  f_V(j_k) f_E(i_k,j_k)=1 .
\]
Therefore it must hold that $q_k=-f_V(i_k)$ for each $k$, which implies that $h$ is a negative feedback function.    

Conversely, if $f_V$ is such that $g\equiv 1$ on $E$ and $h$ is a negative feedback function, 
then $q_k=-f_V(i_k)$.  By the same argument as above, 
$q_k  f_V(j_k) f_E(i_k,j_k)=1$ for all $k$ by monotonicity.  
Therefore $g\equiv -1$ on $E^C$.  Repeating this for all admissible $f_V$, 
maximality follows.    
\epr

There is a second, slightly more sophisticated way of writing a system (\ref{system}) as the feedback loop of a 
system (\ref{controlled system}) using an arbitrary set of edges $E$.    
Given any such $E$, define 
$
S(E^c)=\{ i\, | \mbox{ there is some $j$ such that } (i,j)\in E^c\}.
$  
Now enumerate $S(E^c)$ as $\{ i_1,\ldots i_m\}$, and for each $k$ label the set $\{j\, |\, (i_k,j)\in E^c\}$ as $j_{k1},j_{k2},\ldots$.    
Then for each $k,l$, one can replace each appearance of $x_{i_k}$ in $F_{j_{kl}}$ by $u_k$, to form the 
function $g(x,u)$. Then one lets $h(x)=(x_{i_1},\ldots, x_{i_m})$ as above. The closed loop of this system 
(\ref{controlled system}) is also (\ref{system}) as before but with the advantage that there are 
$|S(E^c)|$ inputs, and of course $|S(E^c)|\leq |E^c|$.    

If $E$ is a consistent and {\em maximal} set, then one can make (\ref{controlled system}) into a monotone system as follows.    
By letting $f_V$ be such that $g\equiv 1$ on $E$, we define the order $\leq_{f_V}$ on $\R^n$.    
For every $i_k,j_{kl}$ such that $(i_k,j_{kl}) \in E^C$, it must hold that $f_V(i_k)f_V(j_{kl})f_E(i_k,j_{kl})=-1$.  
Otherwise $E\cup \{(i_k,j_{kl})\}$ would be consistent, thus violating maximality.    
By choosing $q_k=-f_V(i_k)$, equation (\ref{char 2 eq}) is therefore satisfied.  See the proof of 
Proposition~\ref{propnegativefeedback}. Conversely, if the system generated by $E$ using this second algorithm is 
monotone with respect to orthant orders, and if $h$ is a negative function, then it is easy to verify that 
$E$ must be both consistent and maximal.    

Thus the problem of finding $E$ consistent and such that $P(E_-)=\norma{S(E_-)}=\norma{S(E^C)}$ is smallest, 
when restricted to those sets that are maximal and consistent (this does not change the minimum $\norma{S(E^C)}$), is equivalent to the 
following problem: decompose (\ref{system}) into the negative feedback loop of an orthant monotone control system, 
using the second algorithm above, and using as few inputs as possible.    
This produces the following problem formulation.

\begin{problem}[Directed Labeling Problem($DLP$)]\label{problem2}:

\noindent
An instance of this problem is $(G,h)$ where $G=(V,E)$ is a directed graph and 
$h\colon E\rightarrow\{0,1\}$.
A valid solution is a vertex labeling function $f\colon V\rightarrow\{0,1\}$.
Define an edge $(u,v)\in E$ to be consistent iff
$h(u,v)\equiv (f(u)+f(v)) \pmod{2}$.
The objective is then to find a valid solution {\em minimizing} $|g(E-F)|$ where
$g(C)=\{u\in V\mid\exists y\in V,(u,y)\in C \}$ for any
$C\subseteq E$ and 
$F$ is the set of consistent edges.
\end{problem}

\subsection{Summary of Key Concepts and Results in Approximation Algorithms} 
\label{approx-section}

For any $\gamma\geq 1$ (resp. $\gamma\leq 1$), 
a \emph{$\gamma$-approximate solution} (or simply
an $\gamma$-approximation) of a minimization (resp., maximization) problem is a
solution 
with an objective value no larger than $\gamma$ times (resp., no smaller
that $\gamma$ times) the
value of the optimum, and an algorithm achieving such a solution is said to have
an {\em approximation ratio} of $\gamma$.

In~\cite{lred} Papadimitriou and Yannakakis defined 
the class of MAX-SNP optimization problems and a special approximation-preserving reduction,
the so-called \emph{L-reduction}, that can be used to show MAX-SNP-hardness of an optimization problem.
The version of the L-reduction that we provide below is a slightly modified but equivalent version 
that appeared in~\cite{modlred}. 

\begin{Definition}
\label{def:lred}
\emph{\cite{modlred,lred}} 
Given two optimization problems $\Pi$ and $\Pi'$, we say that $\Pi$ $\emph{L-reduces}$ to $\Pi'$ if 
there are three polynomial-time procedures $T_1$,$T_2$, $T_3$ and two constants $a$ and $b>0$ such that the 
following two conditions are satisfied: 
{\bf (1)} 
For any instance $I$ of $\Pi$, algorithm $T_1$ produces an instance $I' = f(I)$ of $\Pi'$ generated from 
$T_1$ such that the optima of $I$ and $I'$, $OPT(I)$ and $OPT(I')$, respectively, satisfy $OPT(I') \le a\cdot OPT(I)$. 
{\bf (2)} 
For any solution of $I'$ with cost $c'$, algorithm $T_2$ produces another solution with a cost 
$c''$ no worse than $c'$, and algorithm 
$T_3$ produces a solution of $I$ of $\Pi$ with cost $c$ (possibly from the solution produced by $T_2$) satisfying 
$\left|c - OPT(I) \right| \le b\cdot \left|c'' - OPT(I') \right|$.   
\end{Definition}

\noindent
An optimization problem is MAX-SNP-\emph{hard} if any problem in MAX-SNP L-reduces to that problem. 
The importance of proving MAX-SNP-hardness 
results comes from a result proved by Arora et al.~\cite{ptas} which shows that, assuming P$\neq$NP, 
for every MAX-SNP-hard minimization (resp., maximization) 
problem there exists a constant $\varepsilon>0$ such that no polynomial time algorithm can achieve 
an approximation ratio better than $1+\varepsilon$ (resp., better than $1-\varepsilon$).

A special case of the ULP problem, namely when $h(e)=1$ for all $e\in E$,
is the well-known MAX-CUT problem.
An instance of this problem is an undirected graph $G=(V,E)$. 
A valid solution is a set $S\subseteq V$. 
The objective is to find a valid solution that {\em maximizes} the number of edges 
$\{u,v\}\in E$ such that $|\{u,v\}\cap S|=1$. 
The MAX-CUT problem is known to be MAX-SNP-hard.
For further details on these topics, the reader is referred to the excellent book by
Vazirani~\cite{Vazirani:2001:book}.

\subsection{Notations and Terminology}
\label{Notations and Terminologies}

The following notations are used in the rest of the paper.
$V(G)$ and $E(G)$ are the vertex set and edge set of graph $G$, respectively,
and $\widehat{G}$ is the underlying undirected graph of a directed graph 
$G$ obtained by ignoring the directions of the edges.
For $S\subseteq V$, $G(S)$ denotes the subgraph of $G$ vertex-induced by $S$,
and $E_{\text{out}}(S)=\{(u,v)\in E(G)\;|\;u\in S\}$ is the set of out-bound edges of vertices in $S$.
$OPT_{P}(I)$ denotes the size of an optimal solution for a problem $P$ with instance $I$.
The length of a circuit $c$ with respect to weight function 
$w\colon E\mapsto {\mathbb{R}}$ is defined as $\sum\limits_{e\in c}w(e)$;
if no weight function is specified, then $w(e)=1$ for all $e\in E$ is assumed.

\section{Theoretical Results}
\label{Our Results}

Our theoretical results are summarized as follows.

\begin{Theorem}\label{main-theorem}

\noindent
{\bf (a)}
For some constant $\varepsilon>0$, it is not possible
to approximate in polynomial time 
the $ULP$ and the $DLP$ problems to within an approximation ratio of $1-\varepsilon$
and $1+\varepsilon$, respectively, 
unless P=NP.

\noindent
{\bf (b)}
For $ULP$, we provide a polynomial time $\alpha$-approximation algorithm 
where $\alpha\approx 0.87856$ is the approximation factor for the MAX-CUT problem obtained 
in~\cite{Goemans:1995:JACM}
via semidefinite programming. 

\noindent
{\bf (c)}
For $DLP$, if $d_{in}^{max}$ denotes the maximum in-degree of any vertex in the graph,
then we give a polynomial-time approximation algorithm 
with an approximation ratio of at most $d_{in}^{max}\cdot O(\log|V|)$.
\end{Theorem}

Our computational results are illustrated in Section~\ref{appl-sec} by an
implementation of the algorithms applied to a $13$-node Drosophila
segmentation network, as well as to a $200^+$ node recently published network
of the Epidermal Growth Factor Receptor pathway.

\begin{remark}
It should be noted that the complexity of $ULP$ becomes tractable if the network is
biased significantly towards excitatory connections. Obviously, if all the edges of the
given graph $G=(V,E)$ are labeled $0$, then it is possible to label the vertices such that
all the edges are consistent. Moreover, given any graph $G$, it is easy to check 
in $O((|V|+|E|)^3)$ time if an optimal solution contains all the edges as consistent
by solving a set of linear equations via Gaussian elimination.
Now, suppose that at most $L$ of the edges of $G$ are labeled $1$.
Then, obviously at most $L$ inconsistent edges exist in any optimal solution. Thus a 
straightforward way to solve the problem is to consider all possible subsets of edges 
in which at most $L$ edges are dropped and checking, for each such subset, if 
there is an optimal solution that contains all the edges as consistent. The total time 
taken is $O(|V|^{2L}.\cdot (|V|+|E|)^3)$, which is a polynomial in $|V|+|E|$ 
if $L$ is a constant.
\end{remark}

\section{Proof of Theorem~\ref{main-theorem}}
\label{proof of main theorem}

This section provides the proof of Theorem~\ref{main-theorem}, broken up into a
series of technical parts.

\subsection{Proof of Theorem~\ref{main-theorem}(a)}

Based on the discussion in Section~\ref{approx-section},
it suffices to show that both these problems are MAX-SNP-hard.
ULP is MAX-SNP-hard since its special case,
the MAX-CUT problem, is MAX-SNP-hard.
To prove MAX-SNP-hardness of DLP, we need the definitions of the
following two problems.

\begin{problem}[Node Deletion Problem with Bipartite Property ($NDBP$)]\label{prob6}:

\noindent
An instance of this problem is an undirected graph $G=(V,E)$.
A valid solution is a vertex set $S\subseteq V$, such that 
$G(V-S)$ is a bipartite graph.
The objective is to find a valid solution {\em minimizing} $|S|$.
\end{problem}

\begin{problem} [Variance of Node Deletion Problem ($VNDP$)]\label{prob7}
An instance of this problem is 
$(G,h)$ where $G=(V,E)$ is a directed graph and 
$h\colon E\rightarrow\{0,1\}$.
A valid solutions is a vertex set $S\subseteq V$ with
the following property: if 
$G_S=(V_S,E_S)$ is the graph 
with $V_S=V$ and $E_S=E-E_{\text{out}}(S)$, then $\widehat{G_S}$ is free of odd length
circuit with respect to weight function $h$.
The objective is to find a valid solution {\em minimizing} $|S|$.
\end{problem}

First, we note that DLP is {\em equivalent} to VNDP. 
If one identifies the solution set $S$ in UNDP with the solution set 
$g(E-F)$ in DLP, then the set of consistent edges 
$F$ in DLP corresponds to the 
$E_S$ in UNDP since every edge $(u,v)\in F$ satisfying
$h(u,v)\equiv (f(u)+f(v)) \pmod{2}$ is equivalent to stating 
that $\widehat{G_S}$ is free of odd length
circuit with respect to weight function $h$.

Thus, to prove the MAX-SNP-hardness of DLP it suffices
to prove that of VNDP.  
NDBP is known to be MAX-SNP-hard~\cite{Lund:1993:ICALP}.
We provide a $L$-reduction from NDBP to VNDP.
For an instance of VNDP with graph $G=(V,E)$, construct an instance 
of DLP with instance $(G',h)$ as follows (note that $G'$ is a digraph):
$
V'=V(G')= V\cup\{A_{u,v},B_{u,v}\;\mid\;\{u,v\}\in E\}
$,
$
E'=E(G')= \{(u,A_{u,v}),{(A_{u,v},B_{u,v})},{(v,B_{u,v})}\;\mid\;\{u,v\}\in E\}
$, and 
$
h(e)=1 \mbox{ for all } e\in E'
$
Now, the following holds:

\noindent
{\bf (1)} If $S$ is a solution to $NDBP$, it is also a solution to the
generated instance of $UNDP$.
The reason is as follows.      
Notice that every odd length (resp., even length) 
circuit ${\cal C}$ in $G$ corresponds to an odd length (resp., even length)
circuit ${\cal C}'$ in $\widehat{G'}$ with respect to the weight function $h$. 
Since $G(V-S)$ is a bipartite graph, it is free of odd length circuits. 
So for each odd length cycle ${\cal C}$ of $G$, there exists $u\in S$ such that
the deletion of all out-bound edges of $u$ in $G'$ breaks its corresponding 
odd length cycle ${\cal C}'$.
      
\noindent
{\bf (2)} If $S'$ is a solution to $UNDP$, then we can 
construct a solution $S$ of $NDBP$ in the following manner:
for each $x\in S'$:
\begin{quotation}
\noindent
if $x=A_{u,v}$, add $u$ to $T$;\\
if $x=B_{u,v}$, add $v$ to $T$;\\
if $x=u$ or $x=v$, add $x$ to $T$.
\end{quotation}
It is now easy to see that 
since the graph $\widehat{G_{S'}}$ is free of odd length circuit with respect
to $h$, $G(V-S)$ has no odd length circuit either.

Hence, we have 
$OPT_{UNDP}(G',h)\leq OPT_{NDBP}(G)$.
Moreover, given a solution $S'$ of $UNDP$, we are able
to generate a solution $S$ of $NDBP$ such that 
\[
||S|-OPT_{NDBP}(G)|\le ||S'|-OPT_{UNDP}(G',h)| .
\]
Thus, our reduction satisfies the Definition~\ref{def:lred} of 
a L-reduction with $a=b=1$.

\subsection{Proof of Theorem~\ref{main-theorem}(b)}
\label{SDP-section}

Our algorithm for ULP uses the semidefinite programming (SDP)
technique used by Goemans and Williamson in~\cite{Goemans:1995:JACM};
hence we use notations and terminologies similar to that 
used in the paper (readers not very familiar with this technique are
also referred to the excellent explanation of this technique in
the book by Vazirani~\cite{Vazirani:2001:book}).
For each vertex $v\in V$, we have 
a real vector $x_v\in\R^{|V|}$ with $||x_v||_2=1$. Then, we can generate from $ULP$
the following vector program (where $\cdot$ denotes the vector inner product):

\begin{center}
\begin{tabular}{|l|} \hline
Solve the following vector program via SDP methods: \\
\hspace{0.25in} {\em maximize} $\;\frac{1}{2}\sum\limits_{h(u,v)=1}{(1-x_{u}\cdot x_{v})}+\frac{1}{2}\sum\limits_{h(u,v)=0}{(1+x_{u}\cdot x_{v})}$\\
\hspace{0.25in} {\em subject to}:\ 
for each $v\in V$: $x_{v}\cdot x_{v}=1$ 
for each $v\in V$: $x_{v}\in \R^{|V|}$. \\
Select a uniformly random vector $r$ in the $|V|$-dimensional unit sphere
and set\\
$
f(v)=\left\{
           \begin{array}{ll}
             0 & \mbox{if $r\cdot x_v\geq 0$} \\
             1 & \mbox{otherwise} \\
           \end{array}
          \right.
$
\\
\hline
\end{tabular}
\end{center}

This proof of the claimed approximation performance 
of the above vector program is obtained by adapting the proof in 
Section~26.5 of~\cite{Vazirani:2001:book} 
for the MAX-2SAT problem 
to deal with fact that, in our problem, 
$a_{ij}=b_{ij}=\frac{1}{2}$ as opposed to a different set of values 
in~\cite{Vazirani:2001:book}. 
Since there are some subtleties in
adapting that proof for readers unfamiliar with this approach, we provide a sketch of
the proof in the appendix.
The procedure can be 
derandomized via methods of conditional probabilities ({\em e.g.}, see~\cite{MR95}).

\subsection{Proof of Theorem~\ref{main-theorem}(c)}
\label{proof of a lemma}

For an instance of $(G,h)$ of $DLP$, construct instance $(G'=(V',E'),h')$ as follows:
\[
V'=V\cup\{C_{u,v}\;\mid\;(u,v)\in E \;\&\; h(u,v)=0\},
\]
\[
E'=\{e\;\mid\; e\in E\;\&\; h(e)=1\}\cup\{(u,C_{u,v}),(C_{u,v},v)\;\mid\;(u,v)\in E\;\&\; h(u,v)=0\},
\]
and
\[
h'(e)=1 \mbox{ for all $e\in E'$}.
\]
Note that every odd (resp., even) length circuit in $G$ with respect to weight function $h$ corresponds to an 
odd (resp., even) length circuit in $G'$ with respect to weight function $h'$, and vice versa. 
Let $F$ is a set of consistent edges in $(G,h)$ with a vertex labeling function $f$.
Now, observe the following:

\noindent
{\bf (1)}
$F'$ is a set of consistent edges in 
$(G',h')$ with a vertex labeling function $f'$ with $f'(x)=f(x)$ for $x\in V'\cap V$ and 
$f'(C_{u,v})=f(u)=f(v)$ for an edge $(u,v)\in F$ with $h(u,v)=0$; thus,
an edge $(u,v)$ in $F$ correspond to an edge $(u,v)$ in $F'$ if $h(u,v)=1$ and correspond to a pair of edges 
$(u,C_{u,v}),(C_{u,v},v)$ in $F'$ if $h(u,v)=0$. 

\noindent
{\bf (2)}
If $(u,v)\in E-F$ is an inconsistent edge in $(G,h)$,
then the edge $(C_{u,v},v)$ in $G'$ can always be made consistent by choosing $f'(C_{u,v})=f(v)$.

Thus, if $F''$ is the set of consistent edges obtained from $F$ following rules {\bf (1)} and 
{\bf (2)} above, then $|g(E'-F'')|=|g(E-F)|$ and thus 
$
OPT_{DLP}(G',h')=OPT_{DLP}(G,h)
$.
Consider the $NDBP$ problem on $\widehat{G'}$. Any solution to $DLP$ on $(G',h')$ with vertex
labeling function $f'$ and set of consistent edges $F'$ 
cannot contain an odd cycle of consistent edges and thus 
provides a solution to $NDBP$ on $\widehat{G'}$ of size $|g(E'-F')|$. Thus,
$
OPT_{NDBP}(\widehat{G'})\leq OPT_{DLP}(G',h')
=OPT_{DLP}(G,h)
$. 
$OPT_{NDBP}(\widehat{G'})$
can be approximated in polynomial time to within an approximation ratio of
$O(\log|V'|)$~\cite{Lund:1993:ICALP}, \IE, we can find a 
solution $S_{NDBP}(\widehat{G'})$ in polynomial time such that 
$
|S_{NDBP}(\widehat{G'})|\leq O(\log|V'|)\cdot OPT_{NDBP}(\widehat{G'})
\leq O(\log|V|)\cdot OPT_{DLP}(G,h)
$.
Now, $S_{DLP}(G,h)=S_{NDBP}(G')\cup\{u\mid\exists v\in S_{NDBP}(G'),(u,v)\in E\}$,
is obviously a solution to $DLP$ on $(G,h)$. 
Remember that $d_{in}^{max}$ denotes the maximum in-degree of any vertex in $G$.
Thus,
$
|S_{DLP}(G,h)|\leq d_{in}^{max}\cdot |S_{NDBP}(G')|
\leq
d_{in}^{max}\cdot 
O(\log|V|)\cdot OPT_{DLP}(G,h)
$.

\section{Two Examples of Applications of the ULP Algorithm}
\label{appl-sec}

We have implemented the SDP-based algorithm for calculating approximate solutions of the 
undirected labeling problem using Matlab, and we illustrate this algorithm with two 
applications to biological systems.  The first application concerns the relatively small-scale 
$13$-variable digraph of a model of the Drosophila segment polarity network.  
The second application involves a digraph with $300+$ variables associated to the human Epidermal Growth 
Factor Receptor (EGFR) signaling network.  
This model was published recently and built using information from $242$ published papers.

\subsection{Drosophila Segment Polarity}

An important part of the development of the early Drosophila (fruit fly) embryo is the differentiation 
of cells into several stripes (or \emph{segments}), each of which eventually gives rise to an identifiable part 
of the body such as the head, the wings, the abdomen, etc.  Each segment then differentiates into a 
posterior and an anterior part, in which case the segment is said to be \emph{polarized}.  (This differentiation 
process continues up to the point when all identifiable tissues of the fruit fly have developed.)  
Differentiation at this level starts with differing concentrations of certain key proteins in the cells; these 
proteins form striped patterns by reacting with each other and by diffusion through the cell membranes.  

\begin{figure}
\center{\includegraphics[height=2.5in]{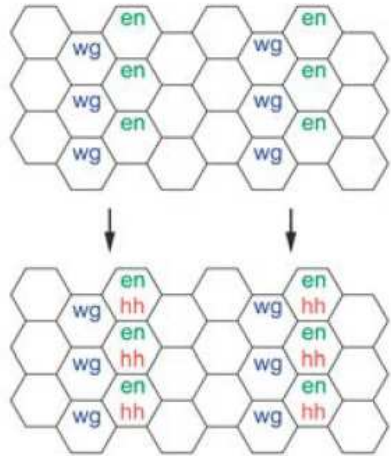}}
\caption{A digram of the Drosophila embryo during early development.  
A part of the segment polarization process is displayed. Courtesy of N. Ingolia and PLoS \cite{Ingolia} }
\label{further figure cells}
\end{figure}

\begin{figure} 
\center{\includegraphics{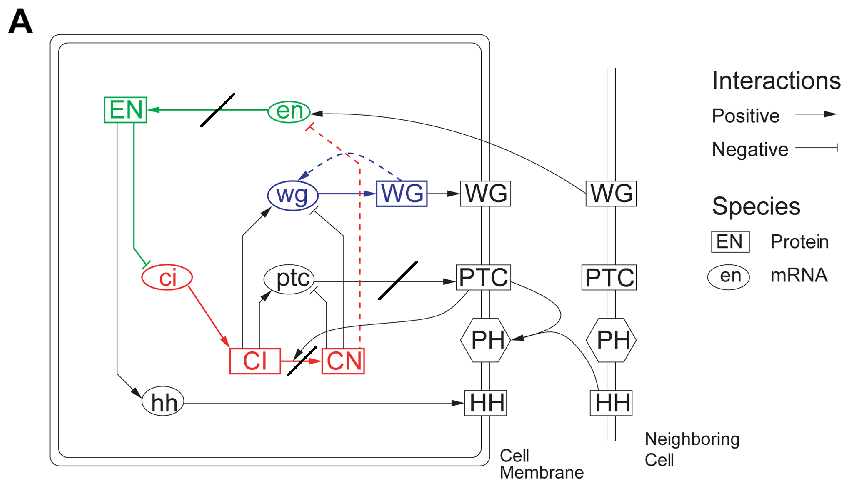}} 
\caption{The network associated to the Drosophila segment polarity, as proposed in \cite{Odell:2000},  
Courtesy of N. Ingolia and PLoS.  The three edges that have been crossed have been 
chosen in order to let the remaining edges form an orthant monotone system.}
\label{further figure Odell}
\end{figure}

A model for the network that is responsible for segment polarity \cite{Odell:2000} is illustrated in 
Figure~\ref{further figure Odell}.  As explained above, this model is best studied when multiple cells are 
present interacting with each other.  But it is interesting at the one-cell level in its own right  --- 
and difficult enough to study that analytic tools seem mostly unavailable.   The arrows with a blunt 
end are interpreted as having a negative sign in our notation.  Furthermore, the concentrations of the 
membrane-bound and inter-cell traveling compounds PTC, PH, HH and WG(membrane) on all cells  
have been identified in the one-cell model (so that, say, HH$\rightarrow$ PH is now in the digraph).  Finally, 
PTC acts on the reaction CI$\rightarrow$ CN itself by promoting it without being itself affected, which 
in our notation means PTC\plus CN \emph{and} PTC\minus CI.  

\paragraph{The Implementation} 

The Matlab implementation of the algorithm on this digraph with $13$ nodes and $20$ edges produced several 
partitions with as many as $17$ consistent edges.  One of these possible partitions simply consists of placing the three nodes 
ci, CI and CN in one set and all other nodes in the other set, whereby the only inconsistent edges 
are CL\plus wg, CL\plus ptc, and PTC\plus CN.  But note that it is desirable for the resulting open 
loop system to have as simple remaining loops as possible after eliminating all inconsistent edges.  
In this case, the remaining directed loops 
\[
\begin{array}{c}
\mbox{EN\minus ci\plus CI\plus CN \minus en\plus EN} \\
\mbox{EN\minus ci\plus CI \plus CN \minus wg\plus WG \plus WG(membrane) \plus en\plus EN}
\end{array}
\]
can still cause difficulties.  

A second partition which generated $17$ consistent edges is that in which EN, hh, CN, and the membrane 
compounds PTC, PH, HH are on one set, and the remaining compounds on the other.  The 
edges cut are ptc\plus PTC, CI\plus CN and en\plus EN, each of which eliminates one or several positive loops.  
By writing the remaining consistent digraph in the form of a cascade, it is easy to see that the only 
loop whatsoever remaining is wg $\leftrightarrow$ WG; this makes the analysis proposed in 
\cite{Enciso:Sontag:DCDS2005} easier.

In this relatively low dimensional case we can prove that in fact $OPT=17$, as the results below will show.   

\begin{Lemma}  \label{further lemma 17}
Any partition of the nodes in the digraph in Figure~\ref{further figure Odell} generates at most $17$ consistent edges.
\end{Lemma}

\bpr
{}From Lemma~\ref{chainlemma}, a simple way to prove this statement is by showing that there are three disjoint cycles 
with odd weighted length
in the network associated to Figure~\ref{further figure Odell} (disjoint in the sense that no 
edge is part of more than one of the cycles).  Such three disjoint cycles exist in this case, and 
they are CI-CN-wg,  CI-ptc-PTC, CN-en-EN-hh-HH-PH-PTC.  
\epr

\subsubsection*{Multiple Copies}

It was mentioned above that the purpose of this network is to create striped patterns of protein 
concentrations along multiple cells.  In this sense, it is most meaningful to consider a 
\emph{coupled} collection of networks as it is given originally in Figures~\ref{further figure cells} 
and~\ref{further figure Odell}.   Consider a row of $k$ cells, each of which has independent concentration variables for 
each of the compounds, and let the cell-to-cell interactions be as in Figure~\ref{further figure Odell} with cyclic 
boundary conditions (that is, the $k$-th cell is coupled with the first in the natural way).  We show 
that the results can be extended in a very similar manner as before.  

Given a partition $\partition$ of the 1-cell network considered above, let $\hpartition$ be the partition 
of the $k$-cell network defined by $\hpartition(\mbox{en}_i):=\partition(\mbox{en})$ for every $i$, etc.  
Thus $\hpartition$ consists of $k$ copies of the partition $\partition$ in a natural way. 

\begin{Lemma} 
Let $\partition$ be a partition of the nodes of the $1$-cell network with $n$ consistent edges.  
Then with respect to the partition $\hpartition$, there are exactly $kn$ consistent edges for the 
$k$-cell coupled model.
\end{Lemma}

\bpr 
Consider the network consisting of $k$ \emph{isolated} copies of the network, that is, $k$ groups of 
nodes each of which is connected exactly as in the 1-cell case.  Under the partition 
$\hpartition$, this network has exactly $kn$ consistent edges.  To arrive to the coupled network, 
it is sufficient to replace all edges of the form 
$(\mbox{HH}_i,\mbox{PH}_i)$ by $(\mbox{HH}_{i+1},\mbox{PH}_i)$ and 
$(\mbox{WG}_i,\mbox{en}_i)$ by $(\mbox{WG}_{i+1},\mbox{en}_i)$, $i=1\ldots k$ (where we identify $k+1$ with $1$).  
Since by definition $\hpartition(\mbox{HH}_{i+1})=\hpartition(\mbox{HH}_i)$ 
and $\hpartition(\mbox{WG}_{i+1})=\hpartition(\mbox{WG}_i)$, 
the consistency of these edges does not change, and the number of consistent edges therefore remains constant. 
\epr

In particular, OPT$\geq 17k$ for the coupled system.  The following result will establish an upper bound for 
OPT.

\begin{figure}
\center{\includegraphics{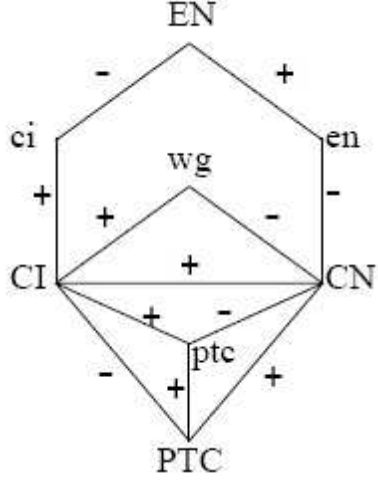}}
\caption{A sub-digraph of the network in Figure~\ref{further figure Odell}, using the notation defined in 
the previous sections.  Note that this sub-digraph doesn't include any of the two 
edges (WGmem,en) and (HH,PH), which connect the networks of different cells in 
Figure~\ref{further figure Odell};  this will be important in the proof of Lemma~\ref{further lemma 17k}.}
\label{further figure planar}
\end{figure}

\begin{Lemma} \label{further lemma 17k}
Any partition of the nodes in the digraph in the $k$-cell coupled network generates at most $17k$ consistent edges.
\end{Lemma}

\bpr
Consider the signed graph in Figure~\ref{further figure planar}, which is a sub-digraph of the network associated 
to Figure~\ref{further figure Odell}.  Since the inter-cell edges (WGmem,en) and (HH,PH) are not in 
this graph, it follows that there are $k$ identical copies of it in the $k$-cell model.  
If it is shown that at least three edges need to be cut in each of these $k$ sub-digraphs, 
the result follows immediately.

Consider the negative cycle ci-CI-wg-CN-en-EN, which must contain at least one inconsistent edge 
for any given partition.  The remaining edges of the subgraph form a tetrahedron with four 
negative parity triangles, which cannot all be cut by eliminating any single edge.  
If follows that no two edges can eliminate all negative parity cycles in this signed graph, and 
that therefore $20k-3k=17k$ is an upper bound for the number of consistent edges in the $k$-cell network. 
\epr

\begin{Corollary}
For the $k$-cell linearly coupled network described in Figure~\ref{further figure Odell}, it holds OPT=$17k$.
\end{Corollary}

\bpr 
Follows from the previous two results.
\epr

\subsection{EGFR Signaling}

The protein called \emph{epidermal growth factor} is frequently stored in epithelial tissues such as skin, and 
it is released when rapid cell division is needed (for instance, it is mechanically triggered 
after an injury).  Its function is to bind to a receptor on the membrane of the cells, aptly called 
the \emph{epidermal growth factor receptor}.  The EGFR, on the inner side of the membrane, has the 
appearance of a scaffold with dozens of docks to bind with numerous agents, and it starts a reaction 
of vast proportions at the cell level that ultimately induces cell division.  

In their May $2005$ paper~\cite{Oda:2005}, Oda et al. integrate the information that has become available about this 
process from multiple sources, and they define a network with $330$ known molecules under $211$ chemical reactions.  
The network itself is available from the supplementary material in SBML format (\emph{Systems Biology Markup 
Language, www.sbml.org}), and will most likely be subject to continuous updates.

\paragraph{The Implementation}

Each reaction in the network classifies the molecules as reactants, products, and/or modifiers (enzymes).  
We imported this information into Matlab using the Systems Biology Toolbox, and constructed a digraph 
$G$ in our notation by letting $\mbox{sign}(i,j)=1$ if there exists a reaction in which $j$ is a product and 
$i$ is either a reactant or a modifier.  We let $\mbox{sign}(i,j)=-1$ if there exists a reaction in 
which $j$ is a reactant, and $i$ is also either a reactant or a modifier.  Similarly $\mbox{sign}(i,j)=0$ 
if the nodes $i,j$ are not simultaneously involved in any given reaction, and $\mbox{sign}(i,j)$ is 
undefined (NaN) if the first two conditions above are both satisfied. 

An undefined edge can be thought of as an edge that is \emph{both} positive and negative, 
and it can be dealt with, given an arbitrary partition, by deleting exactly one of the two signed 
edges so that the remaining edge is consistent.  Thus, in practice, one can consider 
undefined edges as edges with sign $0$, and simply add the number of undefined edges to the number of inconsistent 
edges in the end of each procedure, in order to form the total number of inputs.  
This is the approach followed here; there are exactly $7$ such entries in the digraph $G$.  

\paragraph{The Results}

After running the algorithm $100$ times for this problem, and choosing that partition which produced the 
highest number of consistent edges, the induced consistent set contained 
$633$ out of $852$ edges (ignoring the edges on the diagonal and the $7$ undefined edges).   
See the supplementary material for the relevant Matlab functions that
carry out this algorithm. A procedure analogous to that carried out
for system (~\ref{testosterone}) allows to decompose the system as the
feedback loop of a controlled monotone system using $852-633=219$
inputs. Since the induced consistent set is maximal by definition,
Proposition~\ref{propnegativefeedback} guarantees that the function
$h$ is a negative feedback.

Contrary to the previous application, many of the reactions involve several reactants and products in a 
single reaction.  This induces a denser amount of negative and positive edges: even though there are $211$ 
reactions, there are $852$ (directed) edges in the $330\times 330$ graph $G$. It is very likely that this 
substantially decreases OPT for this system.  

The approximation ratio of the SDP algorithm is guaranteed to be at least
$0.87$ for some $r$, which gives the estimate
OPT$\leq\approx633/0.87\approx728$ (valid to the extent that $r$ has sampled
the right areas of the $330$-dimensional sphere, but reasonably accurate in
practice). 

One procedure that can be carried out to lower the number of inputs is a hybrid algorithm 
involving \emph{out-hubs}, that is, nodes with an abnormally high out-degree.  Recall from the description of the 
$DLP$ algorithm that all the out-edges of a node $x_i$ can be potentially cut at the expense of only one input 
$u$, by replacing all the appearances of $x_i$ in $f_j(x)$, $j\not=i$, by $u$.  
We considered the $k$ nodes with the highest out-degrees, and eliminated all the out-edges associated 
to these hubs from the reaction digraph to form the graph $G_1$.    
Then we run the $ULP$ algorithm on $G_1$ to find a partition $\partition$ of the nodes and a 
set of $m$ edges that can be cut to eliminate all remaining negative closed chains.  
Finally, we put back on the digraph those edges that were taken in the first step, and 
which are consistent with respect to the partition $\partition$.  The result 
is a decomposition of the system as the negative feedback loop of a controlled 
monotone system, using at most $k+m$ edges.  

An implementation of this algorithm with $k=60$ yielded a total maximum number of inputs $k+m=137$.  
This is a significant improvement over the $226$ inputs in the original algorithm.  
Clearly, it would be worthwhile to investigate further the problem of designing efficient 
algorithms for the $DLP$ problem to generate improved hybrid algorithmic approaches.
The approximation ratios 
in Theorem~\ref{main-theorem}(c) are not very satisfactory since $d_{in}^{max}$ and $\log |V|$ could be large factors;
hence future research work may be carried out in designing better approximation algorithms.

We conclude with another, more tentative way to drastically reduce the number
of inputs necessary to write this system as the negative closed loop of a
controlled monotone system.  The idea is to make suitable changes of variables
in the original system using the mass conservation laws.  Such changes of
variables are discussed in many places, for example in~\cite{Volpert^3}
and~\cite{Sontag:mono}.  In terms of the associated digraph, the result of the
change of variables is often the elimination of one of the closed chains.  The
simplest target for a suitable change of variables is a set of three nodes
that form part of the same chemical reaction, for instance two reactants and
one product, or one reactant, one product and one modifier.  It is easy to see
that such nodes are connected in the associated digraph by an odd length
triangle of three edges.

In order to estimate the number of inputs that can potentially be eliminated
by suitable changes of variables, we counted pairwise disjoint, odd length
triangles in the digraph of the EGFR network.  Using a greedy algorithm to
find and tag disjoint negative feedback triangles, we found a maximal number
of them in the subgraph associated to each of the $211$ chemical reactions.
Special care was taken so that any two triangles from different reactions were
themselves disjoint.  After carrying out this procedure we found $196$ such
triangles in the EGFR network.  This is a surprisingly high number,
considering that each of these triangles must have been opened in the $ULP$
algorithm implementation above and that therefore each triangle must contain
one of the $226$ edges cut.  To the extent to which most of these triangles
can be eliminated by suitable changes of variables, this can yield a much
lower number of edges to cut, and it could provide a way to thus stress the
underlying structure of the system.

\section{Supplementary Material: MATLAB Implementation Files}

A set of MATLAB programs have been written to implement the algorithms
described in this paper.  They can be accessed from the URL
\verb+http://www.math.rutgers.edu/~sontag/desz_README.html+. 
The appendix contains more details about it.

\newpage

\appendix

\begin{center}
{\bf APPENDIX}
\end{center}

\section{More Details on SDP Algorithm}

In this appendix, we provide details regarding the proof of the SDP algorithm
for Theorem~\ref{main-theorem}(b) described in Section~\ref{SDP-section}.
The proof method is similar to that used in better-known problems.
For simplicity, we do not describe the derandomization methods and provide
a proof for the expected approximation ratio only.
Define the following notations for convenience:
\begin{itemize}
\item
The vertex set $V$ of the graph for ULP is simply $\{1,2,\ldots,|V|\}$; 
\item
$f_{\rm OPT}$ 
is an optimal vertex labeling for $ULP$ with 
$F_{\rm OPT}$ 
being the set of consistent edges;
\item
SDP$_{\rm OPT}$ 
is the maximum value of the objective value of the vector program 

\begin{center}
\begin{tabular}{l} 
{\em maximize} $\frac{1}{2}\sum\limits_{h(u,v)=1}{(1-x_{u}\cdot x_{v})}+\frac{1}{2}\sum\limits_{h(u,v)=0}{(1+x_{u}\cdot x_{v})}$ \\
{\em subject to}: for each $v\in V$: $x_{v}\cdot x_{v}=1$ \\
\hspace{0.65in} for each $v\in V$: $x_{v}\in \R^{|V|}$ \\
\end{tabular}
\end{center}
\end{itemize}
It is easy to see that 
SDP$_{\rm OPT}\geq |F_{\rm OPT}|$ as follows.
For every $v\in V$ if
$f_{\rm OPT}(v)=0$ then set
\[
x_v=(1,\underbrace{0,0,\ldots,0}_{|V|-1|}),
\]
whereas if
$f_{\rm OPT}(v)=1$ then set 
\[
x_v=(-1,\underbrace{0,0,\ldots,0}_{|V|-1|});
\]
this provides a 
solution for the vector program with 
an objective value of precisely 
$|F_{\rm OPT}|$.
Thus, it suffices if we prove our claim on the approximation ratio relative to
SDP$_{\rm OPT}$ 

Next, note that the vector program can indeed be solved by a SDP approach. 
Let $Y\in\R^{|V|\times |V|}$ be an unknown real matrix with $y_{i,j}$ denoting the
$(i,j)^{\rm th}$ element of $Y$.
It is not difficult to see (via Cholesky decomposition for real symmetric matrices)
that the above vector program is equivalent to the following semidefinite programming 
problem: 
\begin{center}
\begin{tabular}{l} 
{\em maximize} $\frac{1}{2}\sum\limits_{h(u,v)=1}{(1-y_{u,v})}+\frac{1}{2}\sum\limits_{h(u,v)=0}{(1+y_{u,v})}$ \\
{\em subject to}: for each $v\in V$: $y_{v,v}=1$ \\
\hspace{0.65in} $Y$ is a positive semidefinite matrix \\
\end{tabular}
\end{center}
Such a problem can be solved
in polynomial time within an additive error of any constant
$\varepsilon>0$ via 
ellipsoid, interior-point or convex-programming methods 
~\cite{Grotschel:1988:GAC,Vaidya:1989:NAM,Alizadeh:1995:SDP,Nesterov:1989:CVP,Nesterov:1994:CVP}.

Let $\theta_{u,v}$ denote the angle between the two vectors $x_u,x_v\in\R^{|V|}$ in an optimal
solution of the vector program. 
Then, using standard trigonometric results, 
\[
{\rm SDP}_{\rm OPT}=
\frac{1}{2}\sum\limits_{h(u,v)=1}{(1-\cos\theta_{u,v})}+\frac{1}{2}\sum\limits_{h(u,v)=0}{(1+\cos\theta_{u,v})}.
\]

Let $W$ be the expected value of the number of consistent edges of ULP after we have 
performed the randomized rounding step, namely the step:

\begin{center}
\begin{tabular}{l} 
select a uniformly random vector $r$ in the $|V|$-dimensional unit sphere; \\
set $f(v)=\left\{
           \begin{array}{ll}
             0 & \mbox{if $r\cdot x_v\geq 0$} \\
             1 & \mbox{otherwise} \\
           \end{array}
          \right.$ 
\\
\end{tabular}
\end{center}

Then, via linearity of expectation, it follows that 
\[
{\rm E}[W]=
\sum\limits_{h(u,v)=1}{\rm Pr}[f(u)\neq f(v)]+\sum\limits_{h(u,v)=0}{\rm Pr}[f(u)=f(v)].
\]
Because the vector $r$ was chosen randomly, it is true that 
\[
{\rm Pr}[f(u)\neq f(v)]=\frac{\theta_{u,v}}{\pi}
\;\text{and}\;
{\rm Pr}[f(u)=f(v)]=1-\frac{\theta_{u,v}}{\pi}.
\]
Thus,
\beqn
\mbox{E}[W]  &=&
\sum\limits_{h(u,v)=1}\frac{\theta_{u,v}}{\pi}+\sum\limits_{h(u,v)=0} \left(1-\frac{\theta_{u,v}}{\pi}\right)\\
&\geq&
\Delta\cdot
\left[
\frac{1}{2}\sum\limits_{h(u,v)=1}{(1-\cos\theta_{u,v})}+\frac{1}{2}\sum\limits_{h(u,v)=0}{(1+\cos\theta_{u,v})}
\right]\\
&=&
\Delta\cdot
\mbox{SDP}_{\rm OPT} 
\eeqn
where
\[
\Delta=\min\left\{
\frac{2}{\pi}\min_{0\leq\theta\leq\pi}\frac{\theta}{1-\cos\theta},\,
\min_{0\leq\theta\leq\pi}\frac{2-\frac{2\theta}{\pi}}{1+\cos\theta}
\right\}
\]
can be shown to satisfy $\Delta>0.87856$ using elementary calculus.

\subsection{Proof of Lemma~\protect{\ref{chainlemma}}}
\bpr
Suppose that the system is monotone with respect to $\leq_{f_V}$, that is, 
\[
f_V(i)f_V(j) f_E(i,j)=1 \text{for all} i,j,\ i\not=j.
\]
(by Lemma~\ref{characterization2}).    
Let $V(G)=A\cup B$, where $i\in A$ if $f_V(i)=1$, and $i\in B$ otherwise.    
Note that by hypothesis $f_E(i,j)=1$ if $x_i,x_j\in A$ or if $x_i,x_j\in B$.  
Also, $f_E(i,j)=-1$ if $x_i\in A$, $x_j\in B$ or vice versa.  
Noting that every closed chain in $G$ must cross an even number of times between $A$ and $B$, 
it follows that every closed chain has parity $1$.   

Conversely, let all closed chains in $G$ have parity $1$.    
We define a function $f_V$ as follows:  consider the partition of $V(G)$ induced by letting $i\sim j$ 
if there exists an undirected open chain joining $i$ and $j$.   
Pick a representative $i_k$ of every equivalence class, and define $f_V(i_k)=1$, $k=1,\ldots, K$.    
Next, given an arbitrary vertex $i$ and the representative $i_k$ of its connected component, 
define $f_V(i)$ as the parity ($+1$ of  $-1$)  of any undirected open chain joining $i_k$ with $i$.    
To see that this function is well defined, note that any two chains joining $i$ and $j$ can be put 
together into a closed chain from $i_k$ to itself, which has parity $1$ by hypothesis. 
Thus the parity of both open chains must be the same.    

Let now $i,j$ be arbitrary different vertices.    
If $\partial F_j / \partial x_i \equiv  0$, then (\ref{char 1 eq}) is satisfied for $i,j$; 
otherwise there is an edge joining $i$ with $j$.    By construction of the ``potential'' function $f_V$, 
it holds that if $f_V(i)=f_V(j)$ then $f_E(i,j)=1$, i.e.,\ $\partial F_j / \partial x_i \geq 0$, and so (\ref{char 1 eq}) holds as well.    
If $f_V(i)\not=f_V(j)$, then $f_E(i,j)=-1$, i.e.\ $\partial F_j / \partial x_i \leq 0$.    
In that case (\ref{char 1 eq}) also holds, and the proof is complete.    
\epr

\section{Supplementary Material: MATLAB Implementation Files (more details)}

A set of MATLAB programs have been written to implement the algorithms
described in this paper.  They can be accessed from the following URL:

\begin{center}
\verb+http://www.math.rutgers.edu/~sontag/desz_README.html+
\end{center}

The files in this directory are MATLAB functions and scripts in .m format.
They can be opened using any text editor, and each contains descriptions
regarding its purpose and use.  Two useful packages to be used when running
these functions are:
\begin{enumerate}
\item
    The Systems Biology Toolbox for MATLAB, which allows for networks in SBML
    format to be imported into the MATLAB environment.  This toolbox also
    allows for processing of the MATLAB structures as well as the creation of
    SBML format files from MATLAB structures.  It can be downloaded at
    \verb+http://sbml.org/+.

\item
    The SeDuMi Optimization Toolbox, one of the most popular implementations
    of the SDP algorithm for MATLAB.  It is freely available for download at
    \verb+http://sedumi.mcmaster.ca/+.
\end{enumerate}
The most important functions in this directory are listed below:
\begin{description}
\item[(i)] {\tt ReactionDigraph.m}: this function receives a model in SBML format and
produces the associated reaction digraph associated to the reaction.

\item[(ii)] {\tt RepeatPartition.m}: this function produces a partition p which optimizes
the number of consistent edges of a given signed digraph G.  It implements the
SDP-based ULP algorithm.

\item[(iii)] {\tt DLPtrim.m}: this function implements the hybrid ULP-DLP algorithm
mentioned in the end of the discussion of the SGFR network.

\item[(iv)] {\tt PlunderNTriangle.m}: this function uses a greedy algorithm to eliminate odd
parity, pairwise disjoint triangles from a given subgraph of a signed digraph
G (to be used in connection to the discussion regarding changes of variables to
eliminate inputs in the decomposition).
\end{description}
\end{document}